# Nanomagnetic logic with non-uniform states of clocking

Vito Puliafito,[1] Anna Giordano,[2] Bruno Azzerboni,[1] and Giovanni Finocchio[2,*]

[1]*Department of Engineering, University of Messina, C.da Di Dio, I-98166 Messina, Italy*

[2]*Department. of Mathematical and Computer Sciences, Physical Sciences and Earth Sciences, University of Messina, V.le D'Alcontres 31, I-98166 Messina, Italy*

*\*E-mail: gfinocchio@unime.it*

**Abstract**

Nanomagnetic logic transmits information along a path of nanomagnets. The basic mechanism to drive such a transmission, known as clocking, can be achieved by exploiting the spin-Hall effect, as recently observed in experiments on Ta/CoFeB/MgO multilayers [D. Bhowmik, *et al*., Nat. Nano. 9, 59 (2013)]. This paper shows the fundamental mechanism of the spin-Hall driven clocking by using a full micromagnetic framework and considering two different devices, Ta/CoFeB/MgO and Pt/CoFeB/MgO. The former is used for a direct comparison of the numerical results with the experiments while the latter permits to predict the effect of the Dzyaloshinskii-Moriya interaction (DMI) in the clocking mechanism. Results show that the clocking state is non-uniform and it is characterized by the presence of domains separated by Bloch (Néel) domain walls depending on the absence (presence) of the DMI. Our findings point out that for the design of nanomagnetic logic a full micromagnetic approach is necessary.

## 1. INTRODUCTION

The possibility to modify the magnetization of a magnet by means of spin transfer torques, due to spin-polarized currents [1] or spin-orbit interactions [2], has attracted a lot of interest of the scientific community in the last years [3, 4, 5, 6, 7, 8, 9, 10]. Among the technological applications of those spin-based phenomena, the use of the spin-Hall effect (SHE) [5] to design nanomagnetic logic has been proposed [11] with the clear advantage to avoid the need of an external magnetic field.

Nanomagnetic logic, also known as magnetic quantum-dot cellular automata [12], encodes binary information into the easy magnetization direction of single domain nanomagnets. The information is transmitted between two points via a path of nanomagnets through the excitation of a clocking state (the magnetization is set in a state ready to receive information) and its relaxation driven by a reference nanomagnet that stores the information. The coupling mechanism is the dipolar field between neighbor nanomagnets [13, 14]. The main implementation of a clocking process uses a large enough external magnetic field, applied along the in-plane hard axis of the nanomagnets to align the magnetization along that direction, and then, once the field is removed, the magnetization of each magnet will relax towards one of the two easy axis directions depending on the reference nanomagnet [13]. The key problem of this clocking implementation, i.e. presence of random bits, is overcome involving biaxial nanomagnets [15, 16]. From a technological point of view, nanomagnetic logic offers better performances than the semiconductor counterpart in term of integration density and power dissipation [12], however the need of an external field has represented its main limitation. A step forward to the industrial use of nanomagnetic logic is due to the experimental results published in Ref. 11, where current-driven clocking of three nanomagnets CoFeB with perpendicular anisotropy is demonstrated by means of the SHE [5].

Here, firstly we show micromagnetic simulations of the experimental setup Ta/CoFeB/MgO multilayer of Ref. 11. The key result is the finding of a non-uniform clocking state characterized by strip domains separated by Bloch domain walls. The critical clocking current computed with spatially uniform parameters (e.g. anisotropy constant, saturation magnetization) is larger than the experimental value, however this difference, due to possible defects at the edge of the samples, is reduced by introducing in the model a non-uniform spatial profile of the perpendicular anisotropy. The second part of the paper predicts the effects of the Dzyaloshinskii-Moriya interaction (DMI) [17, 18] in the dynamical properties in a similar device (Pt is used as heavy metal being the DMI in Ta/CoFeB negligible) [19, 20, 21, 22]. The presence of DMI gives rise to a non-uniform state characterized by two domains separated by a Néel domain wall. In addition, an unstable region, characterized by continual jumps between the two uniform out-of plane states, is also observed for a range of currents and temperatures. Our results are important from a technological point of view indicating that for the design of nanomagnetic logic a full micromagnetic framework is necessary. On the other hand, from a fundamental point of view, we predict two completely different non-uniform magnetic states in presence/absence of DMI driven by SHE in materials with perpendicular anisotropy.

The paper is organized as follows. Section 2 describes the device geometries under investigation and provides the numerical details of the micromagnetic framework. Results and discussions are given in Section 3, while Section 4 is dedicated to the conclusions.

## 2. DEVICE AND MODEL

The device under investigation is sketched in figure 1. It is composed by an extended heavy metal (Ta or Pt) with on top a bilayer CoFeB/MgO. The CoFeB has dimensions of $200 \times 200 \times 1$ nm$^3$ and an out-of-plane easy axis because of the interfacial perpendicular anisotropy due to the electrostatic interaction between Fe and O in the CoFeB/MgO bilayer [23, 24, 25]. In those devices, an in-plane current $J$ flowing in the heavy metal gives rise to the SHE (circles and crosses in figure 1 indicate the separation of spin-up and spin-down electrons due to the spin-dependent scattering) that generates a spin-transfer torque at the interface with CoFeB [5, 6].

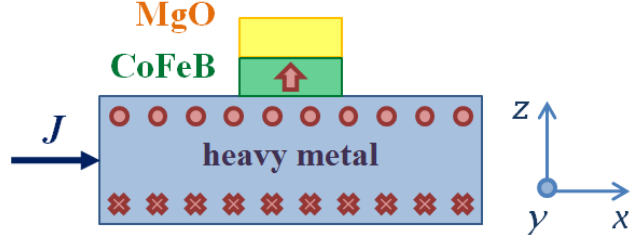

**Figure 1.** Schematics of the device under investigation. The spin-dependent scattering in the heavy metal is responsible for the creation of a spin-current (circles and crosses indicate spin-up and down) and a spin-transfer torque applied to the ferromagnet on top (i.e. CoFeB). The MgO on top of the ferromagnet gives rise to the presence of an interfacial perpendicular anisotropy that drives as easy axis the out-of-plane direction.

The dynamical behavior of the ferromagnet is computed by the numerical solution of the Landau-Lifshitz-Gilbert equation which includes a Slonczewski-type torque to model the STT torque from SHE [6, 8, 26]:

$$\frac{d\mathbf{m}}{\gamma_0 M_S dt} = -\frac{1}{(1+\alpha^2)}\mathbf{m} \times \mathbf{h}_{EFF} - \frac{\alpha}{(1+\alpha^2)}\mathbf{m} \times \mathbf{m} \times \mathbf{h}_{EFF}$$
$$-\frac{d_j}{(1+\alpha^2)\gamma_0 M_S}\mathbf{m} \times \mathbf{m} \times \boldsymbol{\sigma} + \frac{\alpha d_j}{(1+\alpha^2)\gamma_0 M_S}\mathbf{m} \times \boldsymbol{\sigma}$$
(1)

being **m** and **h**$_{EFF}$ the magnetization of the ferromagnet and the effective field, respectively. The **h**$_{EFF}$ includes the micromagnetic contributions from exchange, demagnetizing and interfacial anisotropy fields. External field is not applied in any analysis of this study. In equation (1), $\gamma_0$ is the gyromagnetic ratio, $M_S$ is the saturation magnetization, and $\alpha$ is the Gilbert damping. The coefficient $d_J$ is given by $d_J = \frac{\mu_B \alpha_H}{e M_S t_{FE}} J$, being $\mu_B$ the Bohr magneton, $e$ the electron charge, $t_{FE}$ the thickness of the ferromagnet, and $\alpha_H$ the spin-Hall angle [6]. **σ** is the direction of the spin current in the heavy metal (y-direction in our Cartesian coordinate system, see figure 1) [27, 28]. The effect of the thermal fluctuations is modeled as a stochastic field **h**$_{TH}$ added to the deterministic effective field for each computational cell. The **h**$_{TH}$ is a 3D-vector given by $\mathbf{h}_{TH} = \frac{\xi}{M_S}\sqrt{2\alpha k_B T / \mu_0 \gamma_0 \Delta V M_S \Delta t}$ where $k_B$ is the Boltzmann constant, $\Delta V$ and $\Delta t$ are discretization volume and integration time step, respectively, while $T$ is the temperature. ξ is a 3-D white Gaussian noise with zero mean and unit variance, uncorrelated for each computational cell [29]. For the set of simulations including the interfacial DMI, the **h**$_{EFF}$ also includes the term $\mathbf{h}_{InterfDMI} = -\frac{2D}{\mu_0 M_S}\left[(\nabla \cdot \mathbf{m})\hat{z} - \nabla m_z\right]$ where $\hat{z}$ is the unit-vector of the out-of-plane direction, $m_z$ is the z-component of the normalized magnetization, and $D$ is the DMI parameter. The boundary condition for the exchange interaction is $d\mathbf{m}/dn = -(1/\chi)(\hat{z} \times \mathbf{n}) \times \mathbf{m}$ where $\chi = \frac{2A}{D}$ is a characteristic length in presence of DMI and $A$ is the exchange constant [30, 31].

The simulation parameters used for Ta/CoFeB/MgO are [23, 24, 25]: $A = 2.0 \times 10^{11}$ J/m, perpendicular anisotropy constant $K_U = 6.0 \times 10^5$ J/m$^3$, saturation magnetization $M_S = 800$ kA/m, damping $\alpha = 0.03$, spin-Hall angle $\alpha_H = -0.15$, and $D=0$ mJ/m$^2$. The simulation parameters used for Pt/CoFeB/MgO are the same of the first device, except for $D = 0.5$ mJ/m$^2$ and the spin-Hall angle $\alpha_H = 0.08$ [22, 32]. A squared mesh of $50 \times 50 \times 1$ cells of dimensions $4 \times 4 \times 1$ nm$^3$ is used to discretize the magnet (the exchange length is $l_{exc} = \sqrt{2A/\mu_0 M_S^2} = 7.05$ nm). All the simulations were carried out by considering the positive out of plane magnetization as initial state and positive current density. Quantitative similar results were also achieved by considering negative currents.

## 3. RESULTS AND DISCUSSION

The results on Ta/CoFeB/MgO are summarized in the phase diagram $J$-$T$ of figure 2(a). The two regions, out-of-plane (OOP) and multi-domain/in-plane (MDIP), identify a different response of the magnetization as a function of the current and temperature. In the former, the current is not large enough to modify the initial magnetic state. On the other hand, in the MIDP region the magnetization of CoFeB changes from its uniform out-of-plane state to a multi-domain state characterized by an average $z$-component of the magnetization near zero. The MIDP is the clocking state driven by the SHE.

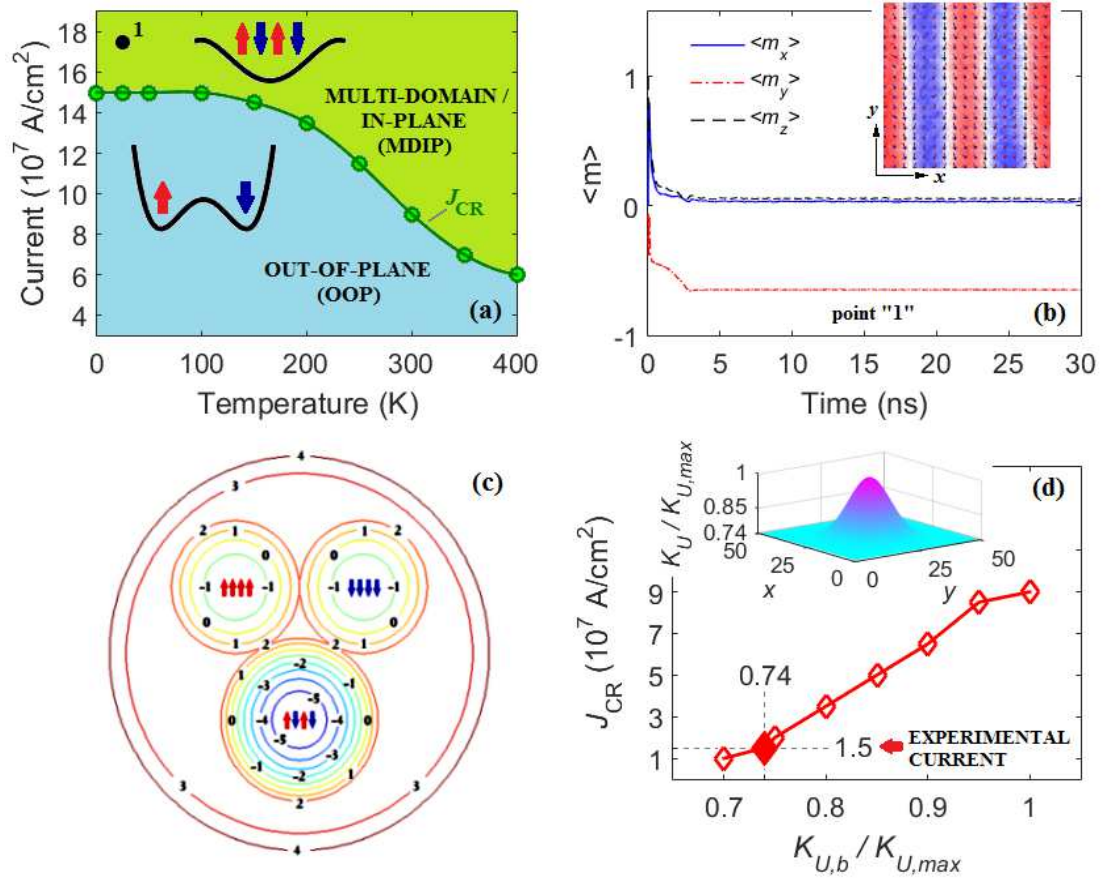

**Figure 2.** (a) Current-temperature phase diagram for the magnetization state of Ta/CoFeB. The insets show the energy landscapes for $T$=0K and $J$=0 A/cm$^2$ (inset in the OOP region) and for $T$=0K and $J > J_{CR}$ (inset in the MDIP region). (b) Time plot of the components of the magnetization of CoFeB in the condition indicated by point "1" of figure (a) ($J$=17.5x10$^7$ A/cm$^2$, $T$=25K) with the corresponding magnetization configuration in the inset. (c) A 2D schematic view of the energy landscape for $T > 0$K and $J > J_{CR}$. (d) Critical current to get MDIP state as a function of the ratio between the value of $K_U$ at the border and its maximum value at the center of the geometry. The filled diamond represents the setup in agreement with the experimental value of critical current. In the inset, the Gaussian profile of $K_U$ as defined by equation (2) in the text.

The phase diagram of figure 2(a) was computed as follows. For a fixed temperature and current density, we ran 25 realizations having duration of 100ns. The critical current density that separates the OOP and the MDIP region was identified as the current density at which one of the 25 realizations gave a final non-uniform state. Figure 2(b) shows the time evolution of the three average components of the normalized magnetization for a realization at $T$=25 K and $J$=17.5x10$^7$ A/cm$^2$ (point "1" of figure 2(a)). After a short transient, the magnetization reaches the configuration displayed in the inset of figure 2(b), where strip domains are separated by Bloch domain walls (the Supplementary Video 1 shows the time domain evolution of the magnetization from the uniform to that strip state). Similar configurations are obtained for all the realizations and, also, at other temperatures (see Supplementary Video 2 for a realization at $T$=300K and $J$=10.5x10$^7$ A/cm$^2$). The direction of the in-plane component of the magnetization of the domain wall, determined by the spin-polarization of the current **σ** and by the sign of the spin-Hall angle, is parallel to the domain wall itself setting a Bloch configuration. The $J_{CR}$ which separates the two regions decreases as a function of the temperature. The origin of this behavior can be understood with the following qualitative explanation. At $T$=0K, the energy landscape at $J$=0A/m is characterized by two minima, uniform up and down out-of-plane configurations (see the inset of figure 2(a) in the region OOP). As the current increases the two uniform states become unstable and the system evolves toward a new minimum (see the inset of figure 2(a) in the region MDIP) that is the non-uniform state, similar to the one displayed in figure 2(b), characterized by magnetic strip domains. At $T$>0K, the scenario becomes more complex. At $J$=0A/m, there are still two minima (uniform states), however as the current increases the third minimum (non-uniform state) appears in the energy landscape and when the thermal energy $E_T=k_BT$ is comparable with the energy barrier $E_B$ ($|J|>J_{CR}$) which separates the three minima, the magnetic state can evolve from its actual configuration to a new local minimum. In particular, our simulations show that the non-uniform minimum is a global minimum and the $E_B$ (uniform→ non-uniform) < $E_B$ (non-uniform→ uniform), see figure 2(c). In this context, the system behaves as an absorbing Markov chain where the non-uniform minimum is the absorbing state [33].The MIDP is unstable without current because of the strong perpendicular anisotropy, in fact the quality factor $Q$ of our film, defined as the ratio between the uniaxial magnetic anisotropy and the self magnetostatic energy ($Q = 2K_U / \mu_0 M_S^2 = 1.49$) is larger than one [34, 35]. A direct comparison of our simulation results with the experimental data of Ref. 11 shows that at room temperature the $J_{CR}$=9x10$^7$ A/cm$^2$ is six times larger than the experimental value (1.5x10$^7$ A/cm$^2$). We investigated on such a difference by following two ways. Firstly, we studied the effect of the Oersted field by computing its contribution to the effective field considering a heavy metal of $400\times400\times5$ nm$^3$. The corresponding simulations highlight a reduction of the critical current density, however it continues to be larger than the experimental result, at $T$=300K a $J_{CR}$=6.5x10$^7$ A/cm$^2$ is achieved which is more than 4 times the experimental achievement, considering a temperature of the ferromagnet of 350K the $J_{CR}$ in presence of the Oersted field is $5.5 \times 10^7$ A/cm$^2$, still more than three times larger. As a second argument, we simulated the presence of defects at the samples edges, due to the etching process, that can reduce locally the perpendicular anisotropy constant $K_U$ [36]. With this in mind, we performed a systematic study of the effect of non-uniform spatial distribution of the $K_U$ over the surface of CoFeB following a Gaussian profile (see inset of figure 2(d)) given by:

$$K_U(x,y) = K_{U,b} + (K_{U,\max} - K_{U,b}) \cdot e^{-\frac{(x-x_c)^2}{2\sigma_1^2}} \cdot e^{-\frac{(y-y_c)^2}{2\sigma_2^2}} \qquad (2)$$

where $K_{U,b}$ is the value of the anisotropy constant at the border of the magnet, $K_{U,\max} = 6.0 \times 10^5$ J/m$^3$ is the maximum value of the constant at the center of the geometry, $x_c$ and $y_c$ are the spatial Cartesian coordinates (see reference system in figure 1) of the geometrical center, $\sigma_1$ and $\sigma_2$ are the standard deviations ($2\sigma_1^2 = 2\sigma_2^2 = 30$ in our study). The $J_{CR}$ as a function of the ratio $K_{U,b}/K_{U,\max}$ is displayed in figure 2(d). In order to achieve a critical clocking current $J_{CR}$ of the same order of the experimental value, a ratio $K_{U,b}/K_{U,\max} = 0.74$ is necessary. Similar analyses were performed at different standard deviations ($2\sigma_1^2$ =20, 40, 50) achieving $0.70 < K_{U,b}/K_{U,\max} < 0.80$. Our results, therefore, show that for the computation of the clocking critical current, together with a small contribution of the Oersted field and a possible increasing of the temperature of the ferromagnet, it is still important to take into account the effect of the defects at the edge of the sample. In particular, one can think to appropriately design the spatial profile of the anisotropy to achieve a low clocking current while maintaining the thermal stability of the ferromagnet. On the other hand, we have to report a negligible effect of randomly distributed defects in the ferromagnet. With this regard, in fact, we performed simulations with different random grain distributions of $K_U$ (it varies from the 90% to the 110% around a central value of $K_U = 6.0 \times 10^5$ J/m$^3$). Those distributions were obtained via an algorithm for Voronoi diagrams. Results pointed out no significant changes in the critical current density with respect to the case of uniform distribution of $K_U$.

In the results of Ref. 11, based on macrospin simulations, the clocking state was attributed to the uniform in-plane configuration with the magnetization aligned along the spin-polarization direction $\sigma$, however here it has been demonstrated that the macrospin approximation is not valid for those devices and the need of a full micromagnetic simulation framework is necessary. Indeed, we found that the uniform in-plane state $|<m_y>|>0.9$ (average y-component of the normalized magnetization) is achieved at much larger current values ($J=23.5 \times 10^7$ A/cm$^2$ at $T$=0K and $J=20 \times 10^7$ A/cm$^2$ at $T$=300K).

The second part of this paper presents results on Pt/CoFeB/MgO multilayer with the aim to understand the effect of DMI on the clocking state of nanomagnets used for nanologic applications. We stress the fact that the differences from the simulations showed in figure 2 are the presence of a DMI field ($D$=0.5mJ/m$^2$) and a spin-Hall angle of 0.08. Figure 3(a) displays the phase diagram *J-T*. Compared to the one in figure 2(a), the key difference is the presence of an "*unstable*" region together with the OOP and MDIP regions. Figure 3(b) shows an example of time domain evolution of the magnetization of this region (point "2" $J$=7.25x10$^7$ A/cm$^2$ $T$=300K). It is characterized by telegraph noise with jumps between the two possible out-of-plane configurations +1 / -1, mediated by a non-uniform state where the nucleation of domains and their expansion/compression occur (see Supplementary Video 3 for a time domain spatial evolution of the magnetization related to the point "2" of figure 3(a)). In this diagram, the critical current density that separates the OOP and the unstable state was identified as the current density at which one of the 25 realizations exhibited in the time trace more than one jump between uniform and non-uniform configuration (similar to Fig. 3(b)), while the one that separates the unstable and the MDIP was identified as the current density at which all the 25 realizations exhibited a single jump from uniform to non-uniform state (similar to Fig. 3(c)). From an energetic point of view, in the unstable region the two uniform states are metastable in the sense that the energy barrier between them is comparable with the thermal energy. This state appears because of the presence of the DMI and in particular because of the change in the boundary conditions which introduce a tilting magnetization at the edge that is a center of domain nucleation. Simulations performed with DMI but without the boundary conditions do not show the unstable region in the *J-T* phase diagram (not shown).

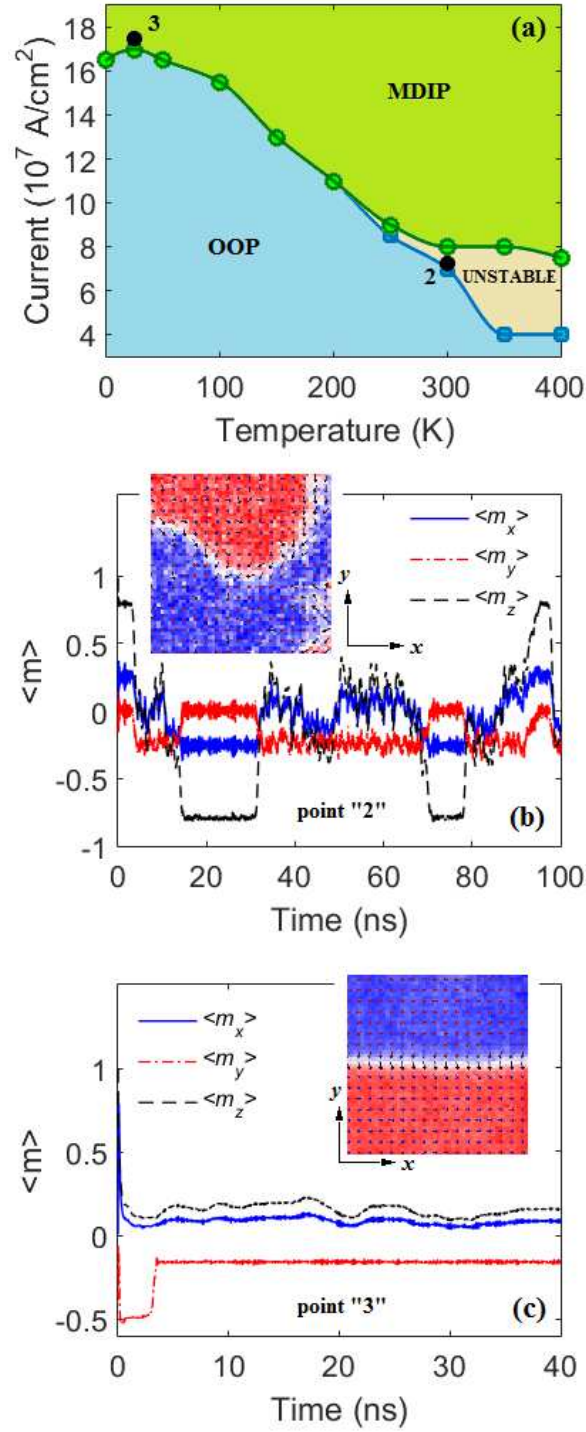

**Figure 3.** (a) Phase diagram for the magnetization of CoFeB as a function of the applied current and the temperature with $D$=0.5mJ/m$^2$. (b),(c) Time plots of the components of the magnetization of CoFeB in the conditions indicated by points "2" ($J$=7.25x10$^7$ A/cm$^2$, $T$=300K) and "3" ($J$=17.5x10$^7$ A/cm$^2$, $T$=25K) of figure (a) with the corresponding obtained magnetization configurations in the insets.

The second key difference between the dynamical behavior of Ta/CoFeB/MgO and Pt/CoFeB/MgO is the magnetic configuration of the clocking state in the MDIP region. Figure 3(c) shows a time domain evolution of the magnetization for $J$=17.5x10$^7$ A/cm$^2$ and $T$=25K (point "3" in figure 3(a)). The non-uniform magnetic state is characterized by two domains separated by a Néel domain wall (the Supplementary Video 4 shows the corresponding time domain evolution of the magnetization configuration from the uniform to that state) and the in-plane component of the domain wall magnetization is perpendicular to the domain wall itself.

These achievements are in agreement with theoretical expectations and experimental results which indicate that the interfacial DMI can stabilize chiral structures, such as Néel domain walls [27, 37]. The DMI parameter which determines the transition from perpendicular to longitudinal to the current domains (from Bloch to Néel domain walls) is $D_{CR}$= 0.3 mJ/m$^2$.

In order to estimate the minimum out-of-plane field to flip the magnetization of the ferromagnet in the clocking state, we performed micromagnetic simulations by adding an external field to the effective field (to simulate the dipolar coupling with a neighbor ferromagnet). Our computations show that the minimum field needed in the unstable region is larger than the one necessary in the MDIP state. For instance, for $T$=300K at $J$=7.25x10$^7$ A/cm$^2$ (point "2" in figure 3(a), unstable region) the minimum field is 30mT while at $J$=10x10$^7$ A/cm$^2$ (MDIP region) it is 10mT. Those estimations show that this unstable region cannot find direct application as clocking state and should be avoided in the working point of real devices for magnetic nanologic.

In order to investigate on the scalability issue, we also studied the properties of a 100×100×1 nm$^3$ ferromagnet, at $T$=0 and 300K, with and without DMI. The key findings and the comparison with the device 200×200×1 nm$^3$ can be summarized into two main points: (i) critical currents are a bit larger in the smaller device (for instance, without DMI and at $T$=300K, $J_{CR}$ is 11x10$^7$ A/cm$^2$, the critical current was 9x10$^7$ A/cm$^2$ in the larger ferromagnet), (ii) as for the case 200×200×1 nm$^3$, the non-uniform magnetization is characterized by perpendicular strips and longitudinal domains at low and high DMI, respectively, however the minimum DMI to nucleate the longitudinal domain is larger in the case 100×100×1 nm$^3$ ($D_{CR}$= 0.6 mJ/m$^2$).

The results reported in the present paper are complementary of other recent reports on SHE-driven magnetization dynamics. In a more recent paper by Bhowmik *et al*. [21], it is reported that in 20 microns wide samples a domain wall, longitudinal to the current, is created and moved by means of SHE and an external in-plane field. Our study shows that in smaller samples the nucleation of longitudinal domain walls is driven by the combined action of SHE and DMI. However, the longitudinal domain wall is unstable without current. In Refs. 38 and 39, the authors focus on magnetization switching in presence of SHE, large DMI (e.g. $D = 1.4$ mJ/m$^2$ in Ref. 38) and an in-plane field. In those papers, the switching process is due to the nucleation of a domain at the edges of the sample that propagates in the whole cross section of the ferromagnet. In our study, we show that for nanologic application the DMI is the key ingredient which can control the clocking state configuration.

## 4. SUMMARY AND CONCLUSIONS

Micromagnetic simulations point out that the clocking states driven by spin-orbit coupling effects such as SHE and DMI are strongly non-uniform and they are qualitatively different from the clocking states driven by the application of an external field (uniform state). While the main effect of the SHE is the destabilization of the uniform state at zero current, the DMI controls the type of the stabilized non-uniform state, perpendicular strips at low D, while longitudinal domain at large D. We have also predicted the presence of unstable state for some range of temperature, D and current density, where the magnetization jumps among the two opposite perpendicular uniform configurations and the non-uniform state. Our results can be used to test in experiments if the DMI plays a significant role in the stabilization of chiral magnetic structures. Our study opens the way to further analyses on nanomagnetic logic.

**ACKNOWLEDGMENTS**

This work was supported by project PRIN2010ECA8P3 from Italian MIUR. The authors thank D. Bhowmik for the useful discussions and for sharing part of his results before the publication, Domenico Romolo for the graphic support, and Giulio Siracusano for the implementation of the algorithm for the Voronoi diagrams.